\documentclass[twocolumn,twoside]{revtex4}
\usepackage{graphicx}
\usepackage{fancyhdr}
\newcommand{\beq}{\begin{equation}}
\newcommand{\eeq}{\end{equation}}
\newcommand{\bea}{\begin{eqnarray}}
\newcommand{\eea}{\end{eqnarray}}

\begin{document}

\title{Impact of Non-perturbative QCD on {\bf CP} Violation in Many-Body Final States \\ of Flavor Transitions}

%

\author{I.I. Bigi\\
Department of Physics, University of Notre Dame du Lac, Notre Dame, IN 46556, USA}
\affiliation{email address: ibigi@nd.edu}

\begin{abstract}
The title of my talk pointed out central statements: the impact of non-perturbative QCD on 
{\bf CP} asymmetries in many-body FS in charm \& beauty hadrons. For practical reasons one measures first 
{\bf CP} violation in two-body final states of heavy flavor hadrons. However, those are 
small parts of charm hadrons and tiny ones for beauty hadrons; therefore one has to probe {\bf CP} asymmetries in three- \& four-body final states. 
Thus the transitions to the many-body FS basically give information about the underlying dynamics. The impact of non-perturbative QCD on {\bf CP} asymmetries 
in many-body FS shows that -- in principle; it is a true challenge even in a semi-quantitative way. One needs correlations with other transitions. 
That is my strategy; however, I have to discuss the tactics on the same level like using consistent parameterization of the CKM matrix. 
Our community has entered a novel era: direct {\bf CP} violation 
has been found in $D^0 \to h^+h^-$ decays \cite{CHARM}. Finally I give short comments about the possible impact of New Dynamics on direct 
{\bf CP} violation in $K_L \to 2 \pi$ and probe {\bf CP} asymmetry in $J/\psi \to \bar \Lambda \Lambda$ transitions. 

\end{abstract}

\maketitle

\thispagestyle{fancy}


\section{Gods = Symmetries speak in Riddles}
I see a connection of the `Gods' with symmetries: not all symmetries are the same; `local' vs. `global', `broken' vs. `unbroken'. 
To describe data one first use models and then model-independent analyses -- indeed true progress. However, the best fitted analyses often do not give the best 
information about the underlying dynamics. Of course, data are the referees -- in the end! We need true collaborations of experimenters and theorists with 
correlations with other transitions -- and `judgments'. The goal for the first quarter of this century (\& this conference): 
establish the existence of New Dynamics (ND) and their features. The tools are: (a) probe many-body non-leptonic final states and 
(b) use collaboration with members of HEP vs. Hadrodynamics from different `cultures'.

\section{Short comments}
Due to the limit of four pages I give only short comments.
\begin{itemize}
\item
For weak decays of $H_Q$ one can use "kinetic scheme" or "potential-subtracted scheme". However, the PDG2018 review basically ignores these schemes, while focus 
on `1S scheme' claiming it gives the same information about the underlying dynamics. However, I quite disagree; the `1S scheme' is not well defined on the 
non-perturbative level!

\item
Wolfenstein's parameterization was very smart \& used all the time. 
The SM with 3 families of quarks describes the CKM matrix with 4 parameters: $\lambda \simeq 0.223$ plus $A$, $\rho$ \& $\eta$ $\sim {\cal O}(1)$. Fitting the data one gets 
$A \simeq 0.84$, but also $\eta \simeq 0.35$ \& $\rho \simeq 0.14$; there is no real control over systematic uncertainties. Furthermore reviews show uncertainties 
${\cal O}(\lambda ^4)$. Now we have gotten a {\em consistent} parameterization \cite{AHN}.

\item
Drawing diagrams is easy, but understanding the underlying dynamics is another thing. One example: re-scattering of 
$\pi \pi \rightleftharpoons  \bar K K$ due to non-perturbative QCD. 

\item
It is crucial to probe {\bf CP} asymmetries in $\Lambda_b^0 $, $\Lambda_c^+$ \& $\Lambda$ decays. 

\end{itemize} 
Basically I have said before -- like at the FPCP2013. 

\section{Broken U- \& V-spin symmetries}
$SU(3)_{\rm flav}$ can be described by 3 $SU(2)$ with I-, U- \& V-spin symmetries. 
Broken U-spin symmetry without V-spin is okay for strong spectroscopy, where (s,d) are combined.  
What about weak decays? In 2005 Lipkin had suggested to subtly use U-spin symmetry \cite{LIPK2}: 
\beq
\Delta = \frac{A_{\rm CP}(B^0 \to K^+\pi^-)}{A_{\rm CP}(B_s^0 \to \pi^+ K^-)} + 
\frac{\Gamma (B^0_s \to \pi^+K^- )}{\Gamma (B^0 \to K^+ \pi^-)} = 0 
\label{LIPKIN}
\eeq
while the LHCb collaboration found in 2018 based on the run-1 \cite{LHCBUSPIN}:
\beq
\Delta_{\rm LHCb} = -\, 0.11 \pm 0.04 \pm 0.03  \; . 
\label{LHCB}
\eeq
While $\Delta_{\rm LHCb}$ is still consistent with zero, it is also consistent with $\sim - 0.1$ as expected for direct 
{\bf CP} violation for two-body final states. 

Correlations of U-spin with V-spin due to re-scattering? PDG2018 shows: $A_{\rm CP}(B^+ \to K^+ \eta) = - 0.37 \pm 0.08$. 
One should learnt two lessons: 

\noindent
(1) The difference between U- \& V-spins is `fuzzy' in weak transitions. 

\noindent
(2) We have to go {\em well beyond} two-body final states to probe {\bf CP} asymmetries!

\section{CP violation in $D^0$ decays}

We have entered a `novel era': (direct) {\bf CP} violation has been found by LHCb \cite{LHCbGuy}: 
\bea
\nonumber 
\Delta A_{\bf CP} &\equiv & A_{\bf CP}(D^0 \to K^+K^-) - A_{\bf CP}(D^0 \to \pi^+\pi^-) 
\\
&=&  ( - 15.4 \pm 2.9 ) \cdot 10^{-4} \; ; 
\label{GUY}
\eea
it is an important achievement! The next question is: where the LHCb collaboration has to `go' now? To establish indirect {\bf CP} violation 
in $D^0 \to K^+K^-$ (\& $D^0 \to K_S\pi^+\pi^-$) or 
direct {\bf CP} asymmetries in other final states, like to probe Dalitz plots 
$D^{\pm}\to \pi^{\pm}\pi^+\pi^-$/$\pi^{\pm}K^+K^- $ or $D^+_s \to K^+\pi^+\pi^-$/$K^+K^+K^-$.
Which lesson can one learn from that? Obviously we need more data. 


What about double Cabibbo suppressed (DCS) ones?  Two very short comments about the LHCb paper JHEP {\bf 04}(2019)063: 

(a) The `Figure 9(a)' there cannot be the leading source; it is misleading to connect the WA diagram with re-scattering. 

(b) Re-scattering gives connections of $D^+ \to K^+K^+K^-$ with $D^+ \to K^+\pi^+\pi^-$. 
We have to wait for run-3 of LHCb to find {\bf CP} asymmetry there. Non-zero values would show there the impact of ND. 

It is crucial to probe {\bf CP} asymmetries in three- \& four-body final states both of charm \& beauty hadrons; I talk about it in the next Section we have examples 
with non-zero values.

\section{CP asymmetries for three- \& four-body Final States}

Two-body final states of suppressed non-leptonic weak decays are a small part of charm mesons \& tiny ones for beauty mesons. 
It means one need much more information about the underlying dynamics. There is a price for working on the 3- \& 4-body final states, 
but also a prize for the underlying dynamics, namely the existence of ND \& its features. The situations are very different for 
$\Delta S = 1\, \& \, 2$ transitions: the final states are two pions, and they are produced by local operators. 
In particular, when one talks about direct {\bf CP} violation, one needs a weak phase, but also strong re-scattering: 
\bea
\nonumber 
|T(\bar H_Q \to \bar a)|^2 - | T(H_Q \to a )|^2   \propto 
\\
\propto  \sum_{a_j \neq a} T^{\rm resc}_{a_j, a}\; {\rm Im}\; T^*_a \,T_{a_j} \neq 0 \; . 
\label{MANY}
\eea 
To understand the information from the data, one needs several tools like chiral symmetry, dispersion relations \& etc. 
Dalitz plots with $\pi$, $K$, $\eta$ \& $\eta^{\prime}$ probe the underlying dynamics with two observables: 
without angular correlations a plot is flat, while resonances \& thresholds show their impact. We have also broad resonances in the 
0.5 - 3 GeV; scalar ones like $f_0(500)$, $K^*_0(700)$ etc. can{\em not} described with a Breit-Wigner parameterization. 

\subsection{Regional CP asymmetries in $B^{\pm}\to K^{\pm}h^+h^-$ \& $B^{\pm}\to \pi^{\pm} h^+h^-$}

LHCb data from run-1 of CKM suppressed $B^+$ decays show no surprising rates:
\bea
\nonumber
{\rm BR}(B^+ \to K^+\pi^-\pi^+) &=&(5.10 \pm 0.29 ) \cdot 10^{-5}
\label{Pen1}
\\
\nonumber
{\rm BR}(B^+ \to K^+K^-K^+) &=&(3.40 \pm 0.14 ) \cdot 10^{-5} \; . 
\label{Pen2}
\eea
Averaged {\bf CP} asymmetries 
are not surprising \cite{AAIJ}: 
\bea
\nonumber
\Delta A_{\rm CP} (B^+ \to K^+\pi^+ \pi^-) &=&  +\, 0.032 \pm 0.008 \pm 0.004 
\\
\nonumber
\Delta A_{\rm CP} (B^+ \to K^+ K^+ K^-) &=&  - \, 0.043 \pm 0.009 \pm 0.003 
\label{AVEK}
\eea
(I ignore production asymmetry of $\pm 0.007$ with $B^{\pm} \to J/\psi K^{\pm}$ as reference mode.). 
`Regional' ones \cite{AAIJ}:
\bea
\nonumber
\Delta A_{\rm CP} (B^+ \to K^+\pi^+ \pi^-)|_{\rm `regional'} = 
\\
= + \, 0.678 \pm 0.078 \pm 0.032 
\\
\nonumber
\Delta A_{\rm CP} (B^+ \to K^+ K^+ K^-)|_{\rm `regional'}  =  
\\ 
= -\,  0.226 \pm 0.020 \pm 0.004     \; . 
\label{REGK}
\eea
The data of even more CKM suppressed $B^+$ decays show no surprising rates:
\bea
\nonumber
{\rm BR}(B^+ \to \pi^+\pi^-\pi^+) &=& (1.52 \pm 0.14 ) \cdot 10^{-5}
\\
\nonumber
{\rm BR}(B^+ \to \pi^+K^-K^+) &=& (0.50 \pm 0.07 ) \cdot 10^{-5} \; .
\label{Pen4}
\eea
However, both averaged {\bf CP} asymmetries 
\bea
\nonumber
\Delta A_{\rm CP} (B^+ \to \pi^+\pi^+ \pi^-) &=&  +\,  0.117 \pm 0.021 \pm 0.009
\\
\nonumber
\Delta A_{\rm CP} (B^+ \to \pi^+ K^+ K^-) &=&  - \, 0.141 \pm 0.040 \pm 0.018
\eea
and `regional' ones \cite{AAIJ}:
\bea
\nonumber
\Delta A_{\rm CP} (B^+ \to \pi^+\pi^+ \pi^-)|_{\rm `regional'} =
\\
= +\, 0.584 \pm 0.082 \pm 0.027
\\
\nonumber
\Delta A_{\rm CP} (B^+ \to \pi^+ K^+ K^-)|_{\rm `regional'} = 
\\
= - \, 0.648 \pm 0.070 \pm 0.013
\eea 
Of course, re-scattering has large impact. One can describe it in the world of hadrons -- like $\pi \pi \rightleftharpoons  \bar K K$  -- 
or in the world of quarks -- $\bar u u/\bar d d \rightleftharpoons \bar s s $. Furthermore they are connected using the word of `duality'.  
Can one predict that semi-quantitatively? It depends on the situations. In particular, one 
needs `judgment' for the definition of `regional' {\bf CP} asymmetries and best connected with other transitions. There is a good chance that the 
LHCb collaboration will change its definition of `regional' {\bf CP} asymmetries after the analyses the data from run-2. Anyway, it is not easy 
for theorists to wait for the results of these analyses. 

\section{CP asymmetries in beauty \& charm baryons}

{\bf CP} asymmetries have been established in strange, beauty and charm mesons, but so far not in the decays of baryons. Of course, one looks 
for direct {\bf CP} violation.

\subsection{Weak decays of beauty baryons}

At the ICHEP2016 conference the LHCb collaboration had shown the data based on run-1 evidence for {\bf CP} asymmetry 
in $\Lambda_b^0 \to p \pi^-\pi^+\pi^-$. In $pp$ collisions one gets different numbers of $\Lambda_b^0$ vs. $ \bar \Lambda_b^0$ 
due to {\em production} asymmetries. Therefore one focuses first on {\bf T}-odd moments. The LHCb experiment has measured the angle between 
two planes: one is formed by the momenta of $p$ \& $\pi^-_{\rm fast}$, while the other one with the momenta of 
$\pi^+$ \& $\pi^-_{\rm slow}$. It has found evidence for {\bf CP} asymmetry on the level of 3.3 $\sigma$ \cite{LHCBBARCP}. Furthermore, the plot given 
at the ICHEP2016 and the Ref.\cite{LHCBBARCP} shows the strength of `regional' {\bf T} asymmetry around $20 \cdot 10^{-2}$. On the other hand, no evidence has been found 
in $\Lambda_b^0 \to p\pi^-K^+K^-$/$pK^-\pi^+\pi^-$/$pK^-K^+K^-$. One can try to `paint' these situations with tree \& penguin diagram. 
However, one cannot claim to understand the underlying dynamics -- yet. Our community has to wait for the data from run-2 based on 
$pp$ collisions at $\sqrt{s}=$ 13 TeV.

\subsection{Weak decays of charm baryons}

For singly Cabibbo suppressed decays
PDG2018 gives  BR$(\Lambda_c^+ \to p\pi^+\pi^-) = (4.2 \pm 0.4)\cdot 10^{-3}$ \& BR$(\Lambda_c^+ \to pK^+K^-) = (1.0 \pm 0.4)\cdot 10^{-3}$. 
These values will be updated from the run-2 of the LHCb experiment `soon' and later by Belle II. Averaged {\bf CP} asymmetries in these Dalitz plots can  be on 
the order of $10^{-3}$ similar to $D^0$ decays as discussed above, see Eq.(\ref{GUY}), and larger for `regional' ones with run-2.

\section{Impact of New Dynamics on strange hadrons?}

Indirect \& direct {\bf CP} violation has been established in the neutral kaon with Re$(\epsilon^{\prime}/\epsilon_K) = (1.66 \pm 0.23)\cdot 10^{-3}$. 
The `Buras team' has argued that the SM can produce only a sizably smaller value like with a factor of two \cite{BURAS2,BURAS}. 
Present LQCD result is somewhat close to that \cite{LATTICE1}. One can hope that future results will clean out the possible impact of ND on 
direct {\bf CP} asymmetry.  
While I am `biased' about this situation, I have to mention the words of my other colleague Pich \cite{PICH}. 

The next step is to probe {\bf CP} asymmetry in strange {\em baryons} as suggested by my colleague G. Punzi from Pisa: LHCb can measure $J/\psi \to \bar \Lambda \Lambda \to [\bar p \pi^+ ] [p\pi^- ]$ 
in the run-3 with a dedicated trigger \& probe {\bf CP} asymmetry below $10^{-4}$. One would get new lessons about the impact of non-perturbative QCD. 
The real goal is to find {\bf CP} asymmetry in $\Lambda$ decays and even to connect the results of $K_L$ \& $\Lambda$ decays. Of course, it is a tough order. 

\vspace*{4mm}

\section{Lessons for the future}

I said it in the beginning: `Gods speak in riddles: tragic oracles and tragic misunderstanding', see FIG.1.  
In quantum field theories one can see a connection between `Gods' and symmetries where I gave examples: some  
symmetries are perfect, while other are broken. I see an analogy in FIG.1, namely HEP experimenters and HEP \& MEP theorists have to work as a `team'. 
\begin{figure}[h!]
\begin{center}
\includegraphics[width=8cm]{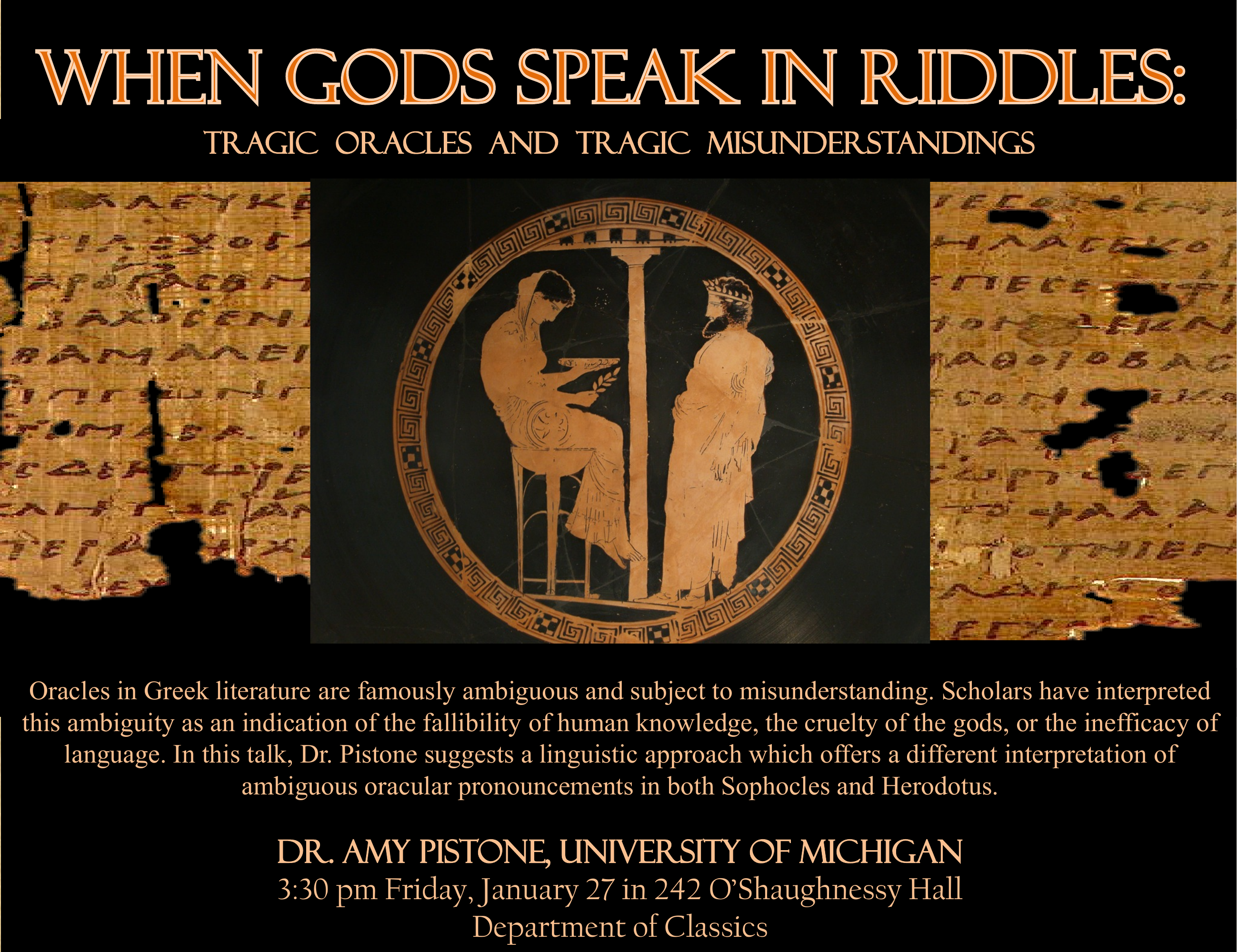}
\end{center}
\caption{When Gods speak in Riddles}
\label{fig:GODS}
\end{figure}

BaBar (\& Belle) were pioneers about weak dynamics \& non-perturbative QCD  and now LHCb \& soon Belle II in a new era.
We need more data, but that is not enough: thinking \& `judgment' about the impact of long-distance QCD from different `cultures': 
`observables = perturb. forces + non-perturb. ones' vs. "observable = long-distance forces + short-distance ones" \cite{BIASED}:
best fitted analyses do not give the best information about the  underlying dynamics; 
{\bf CP} asymmetries in 3- \& 4-body final states is crucial to make progress about ND. 
Challenges between `cultures' of HEP vs. Hadrodynamics like the `masses of current quarks' vs. `pole masses of hadrons' as discussed with details in Ref.\cite{CPBOOK} 
and `soon' in Ref.\cite{GIULIA}. 

Going back to old history: seeing a missile shot by a catapult which had been brought then for the first time, a king from Sparta in the 4th century B.C. cried out: 
`By Heracles, this is the end of man's valor.' Can a theorist see an analogy with computers?

\section{Short comment about $V_{ub}$}

While I had talked about {\bf CP} violation and the impact of non-perturbative QCD, I give very short comments about the situations of $|V_{qb}|$ with $q=c,u$. 
In the present literature the difference between exclusive vs. inclusive data of $|V_{cb}|$ is around $\sim 2 \sigma$ 
\footnote{It was said at FPCP2019 it could be $\sim 3 \sigma$.}. 
However, the discussions about the values $|V_{ub}|$ are $(3 - 4)\, \sigma$ between exclusive vs. inclusive rates with different theoretical tools including LQCD. I had suggested before, 
there could another way to solve this challenge; of course, these loads will go down on the shoulders of our experimenter colleagues. 
Present data about $|V_{ub}|$ could be incomplete in a sizable way: PDG2019 lists branching ratios of 
$B^+ \to l^+\nu \pi^0$/$\eta$/$\eta^{\prime}$ on the level of several$\cdot 10^{-5}$, while $B^+ \to l^+\nu \rho^0$/$\omega$ for $\sim 10^{-4}$ and  
even for BR$(B^+ \to l^+\nu \bar p p) =(5.8^{+2.6}_{-2.3})\cdot 10^{-6}$. Yet PDG2019 has not listed even limits for $B^+ \to l^+ \nu \phi$ or $B^+ \to l^+ \nu K^+K^-$! 
Feynman diagrams give (suppressed) transition  
$B^+ = [\bar b u] \to l^+\nu [\bar u s][\bar s u]$; furthermore  this could be enhanced close to a threshold, which is a subtle item about hadron-quark duality. 
We have the tools to give semi-quantitatively predictions like dispersion relations; `soon' I will work on that.



\vspace*{2mm}

{\bf Acknowledgments:} This work was supported by the NSF PHY-1820860.



\begin{thebibliography}{99}
\bibitem{CHARM}
`Charm' is to ward off `evil spirits' = infinity.
\bibitem{BIASED}
A reader can see that I am biased about the `cultures'.
\end{thebibliography}
\end{document}